\documentclass[letterpaper, 10 pt, conference]{ieeeconf}  

\IEEEoverridecommandlockouts                              

\usepackage{graphics} 
\usepackage{graphicx}
\usepackage{amsmath} 
\usepackage{amssymb}  
\usepackage{gensymb}

\usepackage[utf8]{inputenc} 
\usepackage[T1]{fontenc}

\usepackage{hyperref}

\usepackage{algorithm}
\usepackage{algorithmicx}
\usepackage{algpseudocode}

\usepackage{multirow}
\usepackage[table]{xcolor}

\let\labelindent\relax
\usepackage[inline]{enumitem}

\title{\LARGE \bf
TruPercept: Trust Modelling for Autonomous Vehicle Cooperative Perception from Synthetic Data
}

\author{Braden Hurl$^{1}$, Robin Cohen$^{1}$, Krzysztof Czarnecki$^{1}$, and Steven Waslander$^{2}$
\thanks{$^{1}$Braden Hurl, Robin Cohen, and Krzysztof Czarnecki are with the David R. Cheriton School of Computer Science,
        University of Waterloo.
        {\tt\small [bdhurl, rcohen, k2czarne]@uwaterloo.ca}}%
\thanks{$^{2}$Steven Waslander is with the Institute for Aerospace Studies,
        University of Toronto.
        {\tt\small stevenw@utias.utoronto.ca}}%
}

\begin{document}

\maketitle
\thispagestyle{empty}
\pagestyle{empty}

\begin{abstract}

Inter-vehicle communication for autonomous vehicles (AVs) stands to provide significant benefits in terms of perception robustness. We propose a novel approach for AVs to communicate perceptual observations, tempered by trust modelling of peers providing reports. Based on the accuracy of reported object detections as verified locally, communicated messages can be fused to augment perception performance beyond line of sight and at great distance from the ego vehicle. Also presented is a new synthetic dataset which can be used to test cooperative perception. The TruPercept dataset includes unreliable and malicious behaviour scenarios to experiment with some challenges cooperative perception introduces. The TruPercept runtime and evaluation framework allows modular component replacement to facilitate ablation studies as well as the creation of new trust scenarios we are able to show.

\end{abstract}


\section{Introduction}
\label{intro}
Cooperative Perception, sometimes referred to as collaborative perception or collective perception \cite{collaborativePerception}, is a perception domain which focuses on fusing information from multiple agents. Cooperative perception is grounded in the idea that two or more agents can combine information from their respective viewpoints to increase perceptual range and accuracy and reduce blind spots caused by occlusions from one perspective.

State-of-the-art cooperative perception methods typically assume information received is correct. In real-world situations this assumption may not hold, and malicious attackers could take advantage of models which do not incorporate mechanisms to detect and eliminate false information. Trust modelling is a field of study which aims to identify untrustworthy agents by evaluating the consistency of received information. Trust modelling has been studied for vehicular ad hoc networks (VANETs) (e.g., \cite{MinhasIncentives1}) but never, to the best of our knowledge, with a focus on cooperative perception.

Our aim is to enable the benefits of cooperative perception: extended perceptual range, perception of occluded objects, and increased perceptual accuracy.

The motivation behind using cooperative perception to increase perceptual accuracy, even in situations where the object is at least partially visible to the ego-vehicle, arises from two main ideas:
\begin{enumerate*}
    \item \textit{Ensemble Methods}: aggregate outputs from multiple classifiers to produce scores better than a single classifier \cite{ensemble}. A hypothesis is introduced that using detections from multiple vehicles could improve detection accuracy through similar principles of ensemble learning.
    \item \textit{Viewpoint Diversity}: Different viewpoints may increase detection accuracy of objects in view of the ego-vehicle as a closer, less obstructed, or non-truncated view of an object can increase the probability of a true positive detection.
\end{enumerate*}

The goal is to increase object detection accuracy, even of objects that are beyond the typical sensor range, which will in turn increase vehicle, pedestrian, and cyclist safety. The solution depends on parties being willing to share information and must not degrade performance (i.e., the solution must run in real-time).

Existing work in this area is limited, likely due to the lack of testing data. Current state-of-the-art cooperative perception methods focus on limited test scenarios (e.g., \cite{multiSensorDataFusionPlausibilityChecking}). This makes it difficult to analyze the effect on the entire perceptual pipeline. Cooperative perception without conducting a thorough analysis on the potential implications of malicious agents could lead to danger.

In an endeavor to progress the capabilities and safety of cooperative perception, we offer two primary contributions:
\begin{itemize}
    \item A novel method for integrating trust modelling with cooperative perception into an end-to-end distributed perception model. Experimental results show an increase of up to 5\% Average Precision (AP).
    \item The TruPercept dataset, which is, to the best of our knowledge, the first multi-vehicle perspective perception dataset for AVs gathered in a realistic environment. The TruPercept dataset is not scenario-based but encompasses regular urban driving in a synthetic world. This will allow researchers to benchmark models for improving cooperative perception.
\end{itemize}

These building blocks for effectively understanding trust modelling with perceptual information are essential for expanding robustness of detection methods and exploiting communicated information without leading to overconfidence and susceptibility to simple attacks.

\section{Related Work}
Existing state of the art for cooperative vehicles uses VANETs and various information fusion approaches which are introduced in this section.

Vehicular ad hoc networks (VANETs) are spontaneous networks used for intelligent vehicle communication. VANETs can be comprised solely of intelligent vehicles (IVs), but also commonly include road-side units (RSUs) to extend communication range and a central server for tasks such as moderating agents or information aggregation and evaluation. VANETs can be used to transfer information such as traffic congestion, road conditions, and collision warnings. They will also be the communication medium for cooperative perception. Key issues to address include reliable exchange of messages, even with low latency (e.g., \cite{networkCollisions}, \cite{reladec}) and ensuring the availability of networks for IVs and to optimize RSU locations (\cite{riedl}). Correa et al. \cite{correa} also show how smart infrastructure can enhance perceptual range. Map merging to align views, handling localization between vehicles in case of GPS failures and increasing accuracy have all been examined to provide richer solutions (\cite{KimMultiVehicleSLAM}, \cite{JimenezMultiVehicleSLAM}, \cite{LiCooperativePerception}). Aligning detections shared between vehicles has also been promoted (\cite{allig}). Latency and alignment issues are set aside for now, with our particular model.

Feng et al. exhibit, in their summary of modern object detection and segmentation networks \cite{datasetSummary}, methods which utilize information fusion at various stages in the detection network. Typical fusion methods in 3D object detection (3DOD) fuse image and LiDAR data. Cooperative perception also fuses information from multiple sources; however, from a considerably greater quantity as the sources are nearby vehicles. With cooperative perception the information shared can range from raw image and point cloud data to final object detections or tracks (detections over time).

Arnold et al. \cite{arnold} use images from multiple perspectives to reconstruct 3D objects and discuss how this could be used to better detect relevant 3D objects. Chen et al. \cite{cooper} fuse raw point cloud data which they argue is better than images since there is no need for overlap to perform convolutions. Fusing raw sensor data provides more information to detection networks at the cost of increased bandwidth requirement and latency.

Correa et al. \cite{correa} say perception performance can be enhanced by transporting detection results (late fusion) instead of raw data. They point to two standardized message transportation protocols for late fusion; the Cooperative Awareness Message (CAM) \cite{cam} and the Collective Perception Message (CPM) \cite{cpm}. Hobert et al. \cite{v2xDiscussion} discuss some of the limitations and recommendations for CAMs \cite{cam}. Obst et al. \cite{multiSensorDataFusionPlausibilityChecking} combine tracks from multiple vehicles to increase detection of vehicles that are out of the sensor field of view, which can facilitate detecting and tracking vehicles that enter into view. 

Unfortunately, there is no standard to compare autonomous vehicle cooperative perception methods. Chen et al. \cite{cooper} test their method by concatenating point clouds from the same vehicle of a KITTI scene at two different times. Other works typically use scenario-based test cases and do not include quantitative metrics for large regular driving situations.

Furthermore, there is no guarantee that information received from other vehicles is correct. The cooperative perception methods discussed do not take this into account, which introduces a huge security flaw. A vehicle could disseminate false information unintentionally, for example by using a poorly performing perceptual network, or malicious information could be spread with the intention of causing harm. To prevent this, a central server or authority can be included in VANET system models, often assisting with trust certificates to authenticate trustworthiness (\cite{pomdp}, \cite{rba}). Many trust models include information relayed through vehicles to RSUs then onto the central server; accuracy of information can then be determined if authority-provided ground truth is acquired. Without a central authority (or one unable to obtain ground truth), a trust modelling component then becomes paramount to introduce (e.g., \cite{MessagePropagationAndEvaluation}, \cite{distributedTrustInOpenMultiAgentSystems}).

\section{TruPercept Model}
\label{solution}

Inspired by related work, this section outlines a novel integrative approach to perception and trust for AV perception.

\subsection{Perception}
\label{solPerception}
There are many computer vision models designed for 3DOD in AVs such as \cite{avod} and \cite{fpointnet}, which can be interchanged within the TruPercept system. An object instance, denoted as $\theta$, is detailed with a class (vehicle, pedestrian, cyclist), 3D position relative to on-board GPS ($x,y,z \in \mathbb{R}^3$), 3D bounding box dimensions (width, height, length $\in \mathbb{R}^3$), heading (rotation around up axis $\in \{-\pi,\pi\}$), and a score. The set of all possible detections is denoted $\Theta$ so that $\theta \in \Theta$. The ego-vehicle, the perspective the model is defined from, is denoted as $\alpha$ and a superscript is used to denote agent perspective (i.e., $\Theta^\alpha$ are the local detections output by $\alpha$). Detections and their evaluations will be broadcast to nearby vehicles, which can determine to rebroadcast if the message is within a preset range and time period.

\subsection{Detection Evaluation}
\label{solProposalEval}
Judging trustworthiness of reports received from peers is often calculated by evaluating messages against ground truth data. For 3DOD ground truth data is true information ($\theta$) for each object in range, which, if available, would make perceptual pipelines unnecessary. Instead, each vehicle will evaluate each detection it receives with the information from its perspective. The following function definitions are used:

\begin{itemize}
    \item IoU: $\Theta \times \Theta \to [0,1]$: Returns the Intersection over Union (IoU) of two 3D bounding boxes.
    
    \item $s^v: \Theta \to [0,1]$: Returns the detection score of a detection from a vehicle $v$ which attempts to represent confidence \cite{avod} or probability \cite{bayesOD} of a true detection.
    
    \item $e^v: \Theta \to [0,1] \hspace{1mm} \cup \hspace{1mm} -1$: The evaluation score (from a vehicle $v$) of a detection. Evaluation scores should ideally be probabilistic between 0 and 1, where 0 represents that the detection is incorrect and 1 signifies the vehicle $v$ is 100\% certain the detection is correct.
\end{itemize}

The ego-vehicle, denoted as $\alpha$, is already evaluating whether detections are present or not, within the FoV of on-board sensors, from a local 3DOD network. It follows naturally to evaluate received detections on whether or not they match $\Theta^\alpha$. Firstly, all received detections are filtered by local detection area (70 metres 90\degree FoV for AVOD \cite{avod} default configuration). Next, received detections are matched with $\Theta^\alpha$, producing a set $\mathcal{M}$ of sets of matches $\Theta_m$. A matching set $\Theta_m$ has a detection from $\alpha$ (if matched), followed by all matching received detections. Two detections are considered a match if the IoU is over a threshold parameter $\tau$ and are the same object class (e.g., car, pedestrian, cyclist). Each received detection will be matched to the ego-vehicle detection with which it has the highest IoU. Only one detection per received vehicle detection list is matched with any ego-vehicle detection. Any detection $\theta$ which $\alpha$ receives which matches one of its own detections $\theta'$ will have $e^\alpha$ equal to $s^\alpha(\theta')$. If there is no match with $\theta$, $\alpha$ will set $e^\alpha(\theta)$ to $\eta$, a negative evaluation constant. Experiments are conducted with $\eta$ set to either 0 or -1. For each received detection $\theta$, and the detection of $\alpha$ labelled $\theta'$, where $\theta'$ is the object instance in $\Theta^\alpha$ which has the maximum IoU with $\theta$, the following applies
\begin{equation}
e^\alpha(\theta) =
	\begin{cases} 
      s^\alpha(\theta') & \text{IoU}\left(\theta,\theta'\right) > \tau \\
      \eta & \text{otherwise}
   \end{cases}
\end{equation}

Using only the local detection score to evaluate detections is not sufficient. Objects which are further away, truncated, or occluded may have lower or $\eta$ evaluation scores which would be erroneous evaluations. To solve this problem, another value which encompasses the visibility of the detection is proposed using the following definitions:

\begin{itemize}
    \item $p^v: \Theta \to \mathbb{Z}^\geq$: the number of points from the perspective vehicle $v$ LiDAR point cloud $\mathcal{P}^v$ which are within the boundaries of an object $\theta$.
    
    \item $\Phi^v: \Theta \to [0,1]$: the visibility value of a detection $\theta$ from a vehicle $v$. 1 is completely visible and 0 signifies no knowledge from the current viewpoint.
\end{itemize}

The true visibility of an object $\theta$ from a perspective $v$ is difficult to calculate and will be approximated by $p^v(\theta)$. Let $\gamma_l$ and $\gamma_u$ be the lower and upper $p^v(\theta)$ limits for min/max visibility respectively. Visibility is calculated as
\begin{equation}
\Phi^v(\theta) = \min\left(1, \frac{\max\left(0,p^v(\theta) - \gamma_l\right)}{\gamma_u - \gamma_l}\right)
\end{equation}

If there are no points residing in the 3D bounding box then the object is likely obstructed and the visibility value will be set to 0 (the vehicle is completely unsure as to whether the received detection is correct or not). This follows the assumption that the more LiDAR points that return from an object, the higher the chance the 3DOD algorithm has of detecting it. Thus, the visibility score can be used as a means to estimate the confidence that there is no object for a negative (i.e., unmatched) detection evaluation.

\algnewcommand\algorithmicforeach{\textbf{for each}}
\algdef{S}[FOR]{ForEach}[1]{\algorithmicforeach\ #1\ \algorithmicdo}
\let\oldReturn\Return
\renewcommand{\Return}{\State\oldReturn}



\subsection{Trust}
\label{solTrust}
It can also be beneficial to evaluate the total information flow from a vehicle. For example, if a vehicle is broadcasting malicious information, the vehicle should be identified, so it can be ignored and appropriate measures can be taken. Trust calculation can be done centrally or by each vehicle. Central aggregation introduces a strong system requirement (central server), but is better as vehicles only enter within proximity of each other for short time periods ($\approx 15$ seconds average in the TruPercept data). Trust values are calculated for each object and then vehicle on the central server and then periodically broadcast so vehicles may use the information while integrating detections from other vehicles.

The trust $\mathcal{T}$ for a detection $\theta^\alpha$ will be denoted as $\mathcal{T}(\theta^\alpha)$ and will be aggregated using evaluations from all nearby vehicles $V$ which received the detection $\theta^\alpha$ ($V$ excludes $\alpha$). The evaluation from each vehicle $v \in V$ will be aggregated in proportion to how visible the detection is from each evaluator perspective $v$ (i.e., $\Phi^v(\theta)$). Let the trust function for a detection from the ego-vehicle $\theta^\alpha$ be $\mathcal{T}: \Theta \to [0,1]$.
\begin{equation}\label{tm_eval}
\mathcal{T}(\theta^\alpha) = \frac{\sum\limits_{ v \in V} \Phi^v(\theta^\alpha) \cdot e^v(\theta^\alpha)}{\sum\limits_{v \in V} \Phi^v(\theta^\alpha)}
\end{equation}

The trust value for a vehicle $v$ is calculated on a central server by aggregating the trust feedback from all detections that vehicle has sent in proportion to the local detection score for each detection. Up to this point, we have done everything instantaneously, but trust is calculated over time. Let $\Theta^\alpha_t$ be $\Theta^\alpha$ from a time step $t$. Let $\mathcal{B}^\alpha$ be the set of all detections from $\Theta^\alpha_t$ for every $t \in [-f,0]$, where $f$ is the freshness length. Let the trust function for vehicles be $\mathcal{T}: \mathcal{B} \to [0,1]$. Then the trust for $\alpha$ is
\begin{equation} \label{eqnTv}
\mathcal{T}(\mathcal{B}^\alpha) = \frac{\sum\limits_{ \theta \in \mathcal{B}^\alpha} s^\alpha(\theta) \cdot \mathcal{T}(\theta)}{\sum\limits_{\theta \in \mathcal{B}^\alpha} s^\alpha(\theta)}
\end{equation}

\subsection{Detection Aggregation}
\label{solAddToPerc}

A final aggregation step (at each vehicle) produces a score $\omega$ for every matching set $\Theta_m \in \mathcal{M}$ where $\theta_{mi}$ is the $i$th detection $\theta \in \Theta_m$. There are several inputs to this stage: evaluations $e^v(\theta)$, visibility $\Phi^v(\theta)$, and trust of evaluator $\mathcal{T}(\mathcal{B}^v)$. Let the score function for a final detection be $\omega: \Theta_m \to [0,1]$.

\subsubsection{Weighted Average}
A simple aggregation system which uses a weighted average calculates final scores as
\begin{equation}\label{aggFormula}
\omega^\alpha(\Theta_m) = \frac{\sum\limits_{\theta^v \in \theta_m} \Phi^v(\theta^v) \cdot \mathcal{T}(\mathcal{B}^v) \cdot e^v(\theta^v)}{\sum\limits_{\theta^v \in \Theta_m} \Phi^v(\theta^v) \cdot \mathcal{T}(\mathcal{B}^v)}
\end{equation}
$\eta = 0$ for this method so that $\omega(\Theta_m) \in [0,1]$.

\subsubsection{Additive (Positive and Negative)}
The weighted average method is able to quantify the belief an object is present relative to the trust and visibility of the object. However, it does not account for the quantity of vehicles which perceived the object. It is expected that the higher the ratio of vehicles which perceive a visible object, the likelier the object is to exist. The additive aggregation adds to the final detection score for every vehicle which perceived the object and subtracts for every vehicle from which the object was not perceived (unmatched). $\eta = -1$ to create the positive/negative addition/subtraction mechanism. The aggregation is weighted by the object visibility $\Phi^v(\theta)$ and trustworthiness $\mathcal{T}^v$ of each evaluator $v$. The final score is

\begin{equation}\label{addAggFormula}
\omega^\alpha(\Theta_m) = \sum\limits_{\theta^v \in \Theta_m} \Phi^v(\theta^v) \cdot \mathcal{T}(\mathcal{B}^v) \cdot e^v(\theta^v)
\end{equation}

The resulting value of equation \ref{addAggFormula} is not bounded by 0 or 1; it is later bounded using $\omega^\alpha(\Theta_{m}) = \min\left(1, \max\left(0, \omega^\alpha(\Theta_{m})\right)\right)$. Trust value for the ego-vehicle should be set to 1, or higher than 1 if bias towards the ego-vehicle detections is desired.

\subsection{Plausibility Checker}
Thus far, the message evaluation relies on visibility of objects. This could be taken advantage of if a malicious agent were to insert false detections where there are no points. For example, false detections are inserted half a meter off the ground in front of vehicles where there is open air. No LiDAR points would be registered and the visibility value would be calculated as zero even though the non-object is visible. If there are points behind the object, and no points in front or within the object bounding box, then the existence of the object is false with high confidence. Obst et al. \cite{multiSensorDataFusionPlausibilityChecking} create a mechanism for determining false tracks for cooperative perception and use the term plausibility checker.

The TruPercept system incorporates a novel plausibility checker which performs a frustum cull on the point cloud. The frustum is centered to the object center and extends a quarter of the smallest dimension (width or length) to the top, bottom, left, and right. If more than 10\% of the points are closer than the object center, it is considered plausible.

The plausibility check can be performed in two key spots:
\begin{enumerate*}
    \item During message evaluation it can provide strong negative evaluations instead of not contributing to the evaluation due to low/zero visibility.
    \item After aggregating detections to remove any detections which were erroneously aggregated to scores higher than zero.
\end{enumerate*}

\subsection{TruPercept Summary}
\label{systemSummary}
The TruPercept system has the following detection and decision-making cycle on the ego-vehicle $\alpha$: obtain sensor data; run the detection network to obtain $\Theta^\alpha$; broadcast $\Theta^\alpha$; receive $\Theta^v \hspace{1mm} \forall \hspace{1mm} v \in V$ or stop at timeout $t$; calculate $\mathcal{M}$; calculate and broadcast $p(\theta), e(\theta) \hspace{1mm} \forall \hspace{1mm} \theta \in \Theta^v, v \in V$; receive $p(\theta), e(\theta) \hspace{1mm} \forall \hspace{1mm} \theta \in \Theta^v, v \in V$ or stop at timeout $t$; calculate $\omega(\Theta_m) \hspace{1mm} \forall \hspace{1mm} \Theta_m \in \mathcal{M}$.

The central server receives vehicle broadcasts of $e(\theta)$ and updates $\mathcal{T}(\theta)$ and $\mathcal{T}(\mathcal{B}^v) \hspace{1mm} \forall \hspace{1mm} \theta \in \mathcal{B}^v, v \in V \cup \alpha$ which it then periodically broadcasts.



\section{The TruPercept Dataset}
\label{sec:truperceptData}
Ideally, data would be collected from the real-world, as it would most accurately represent autonomous driving conditions; however, collecting data from the real-world for cooperative perception has resulted only in limited test-case scenarios. To the best of our knowledge, there are no existing publicly available real-world or synthetic cooperative perception datasets for autonomous driving encompassing regular driving. Creating a real-world dataset would require a fleet of vehicles equipped with autonomous sensor systems simultaneously collecting data.

Trust modelling for V2V are often evaluated through the use of simulation software. Chen et al. \cite{pomdp} use OMNET++ \cite{omnet} for V2V communication and SUMO \cite{sumo} for road traffic simulation. However, these simulators do not contain sensory level perceptual information. Simulators such as CARLA \cite{carla} are a viable options as they can obtain synthetic perceptual data and attempt to simulate traffic flow. Alternatively, Grand Theft Auto V (GTA V) is a video game which contains realistic graphics and has been used to generate synthetic data in a diverse environment for AV perception \cite{driveInMatrix}. Recent efforts have even created an in-game LiDAR point cloud generator \cite{gtavlidar}. Hurl et al. \cite{presil} extend the range of \cite{gtavlidar} and improve the accuracy of point clouds for pedestrians. The PreSIL dataset \cite{presil} produces the necessary data for 3DOD for autonomous driving and has been shown to increase performance on state-of-the-art 3DOD networks.

The TruPercept dataset uses the PreSIL generation method \cite{presil} and consists of two series of captures, spanning approximately 300 and 400 seconds respectively. The data is captured at 1 Hz and contains identical data to the PreSIL dataset for the ego-vehicle and each vehicle within a 100 metre radius. A 100 metre radius is used to restrict the total number of vehicles. Ideally, 160m will be used as the LiDAR scanner used for the PreSIL dataset has a max range of 80m so vehicles must be within double the max LiDAR range to have overlapping point coverage areas and thus obtain matching detections. The data is captured over two routes through urban environments.

\section{Qualitative Analysis}

\subsection{Scenario 1: Algorithm Analysis Using a Single-Frame}
\label{baselineScenario}
A scenario was constructed to evaluate qualitatively whether the novel cooperative perception solution can detect a completely occluded object and filter out a malicious detection. The scenario contains the ego-vehicle passing by a parked truck. A pedestrian, occluded from the ego-vehicle perspective, is walking out from behind the truck attempting to illegally cross the road. Another vehicle moving in the opposite direction of the ego-vehicle has a clear line of sight to the pedestrian. A false detection in a dangerous location (less than 10m away, directly in its trajectory, and oncoming) for the ego-vehicle is inserted into the broadcast of the oncoming vehicle. The false detection is given a high detection score (1.0) in an attempt to fool the ego-vehicle.

\begin{figure}[htbp]
  \centering
  \vspace{2.5mm}
  \includegraphics[width=0.2315\textwidth]{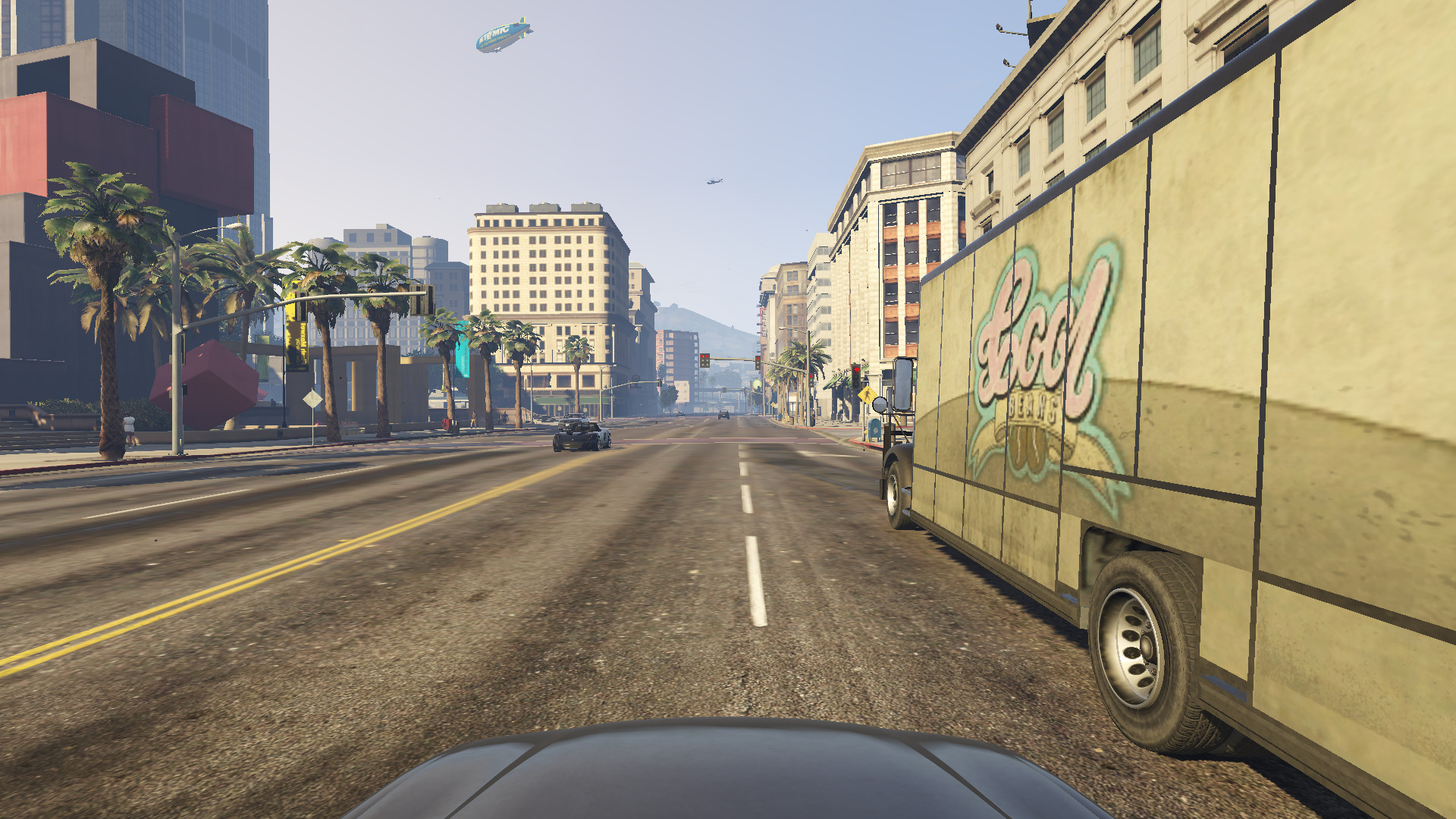}
    \includegraphics[width=0.23\textwidth]{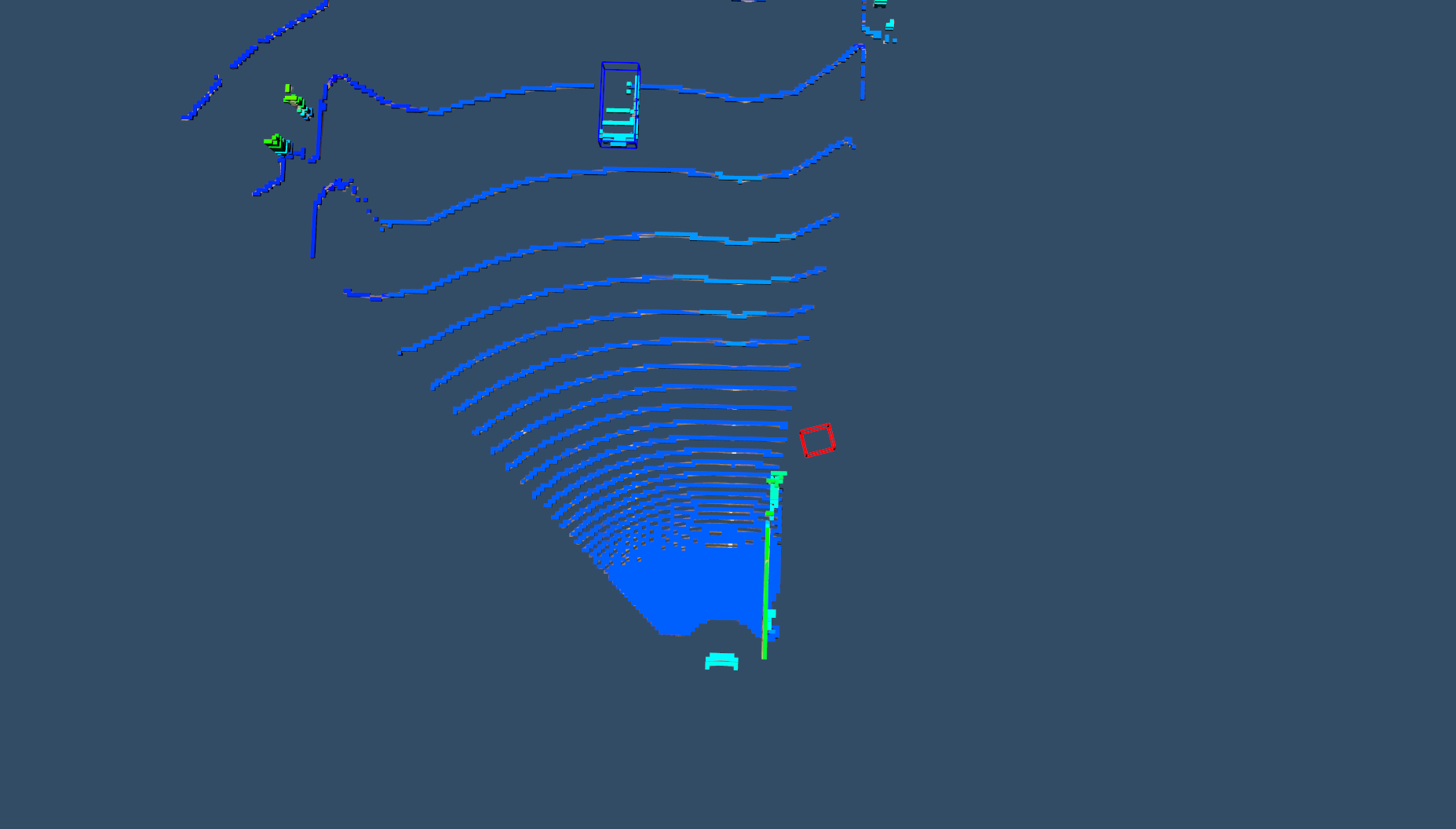}\\
    \vspace{1mm}
    \includegraphics[width=0.23\textwidth]{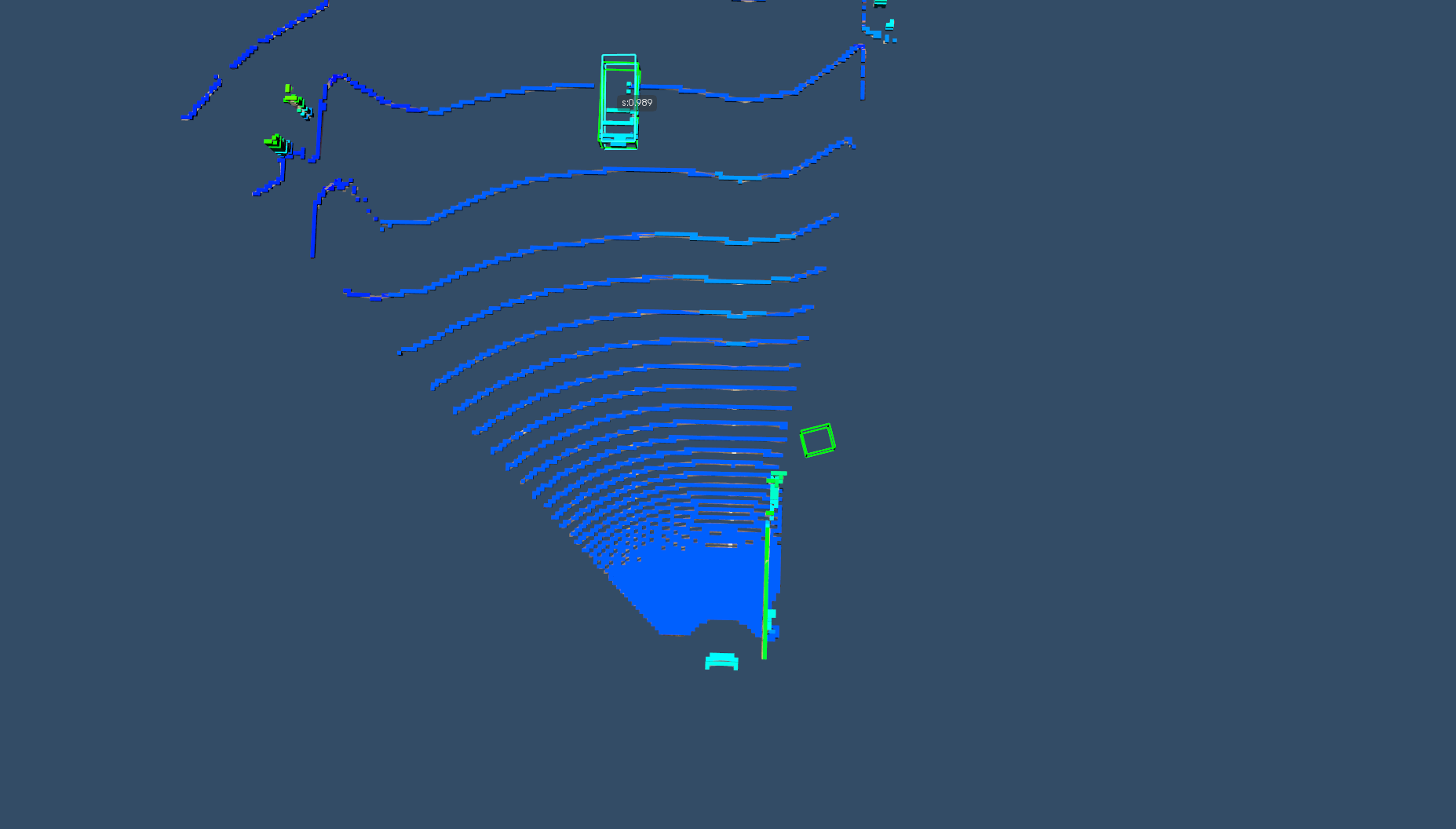}
    \includegraphics[width=0.23\textwidth]{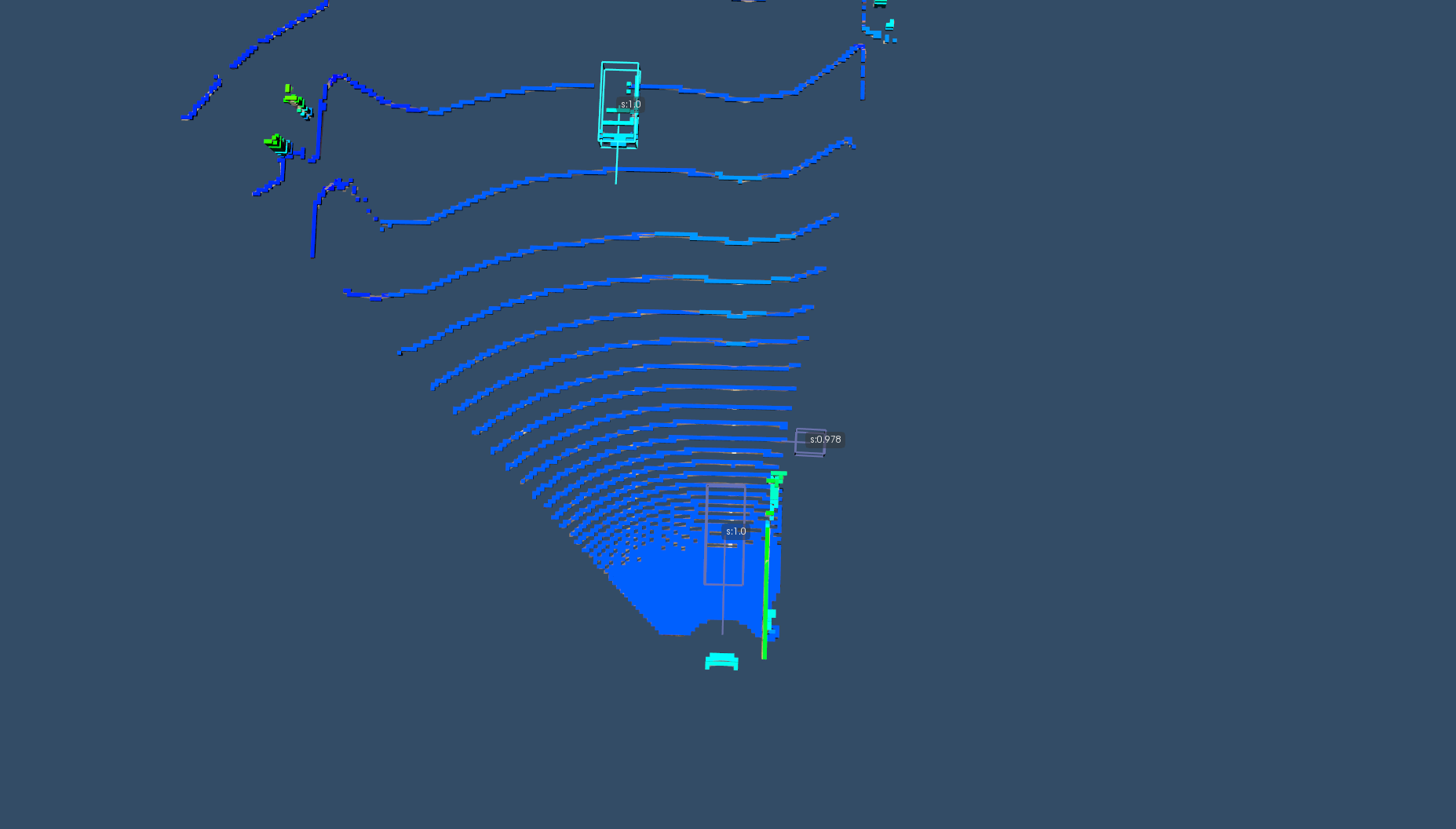}\\
    \vspace{1mm}
    \includegraphics[width=0.23\textwidth]{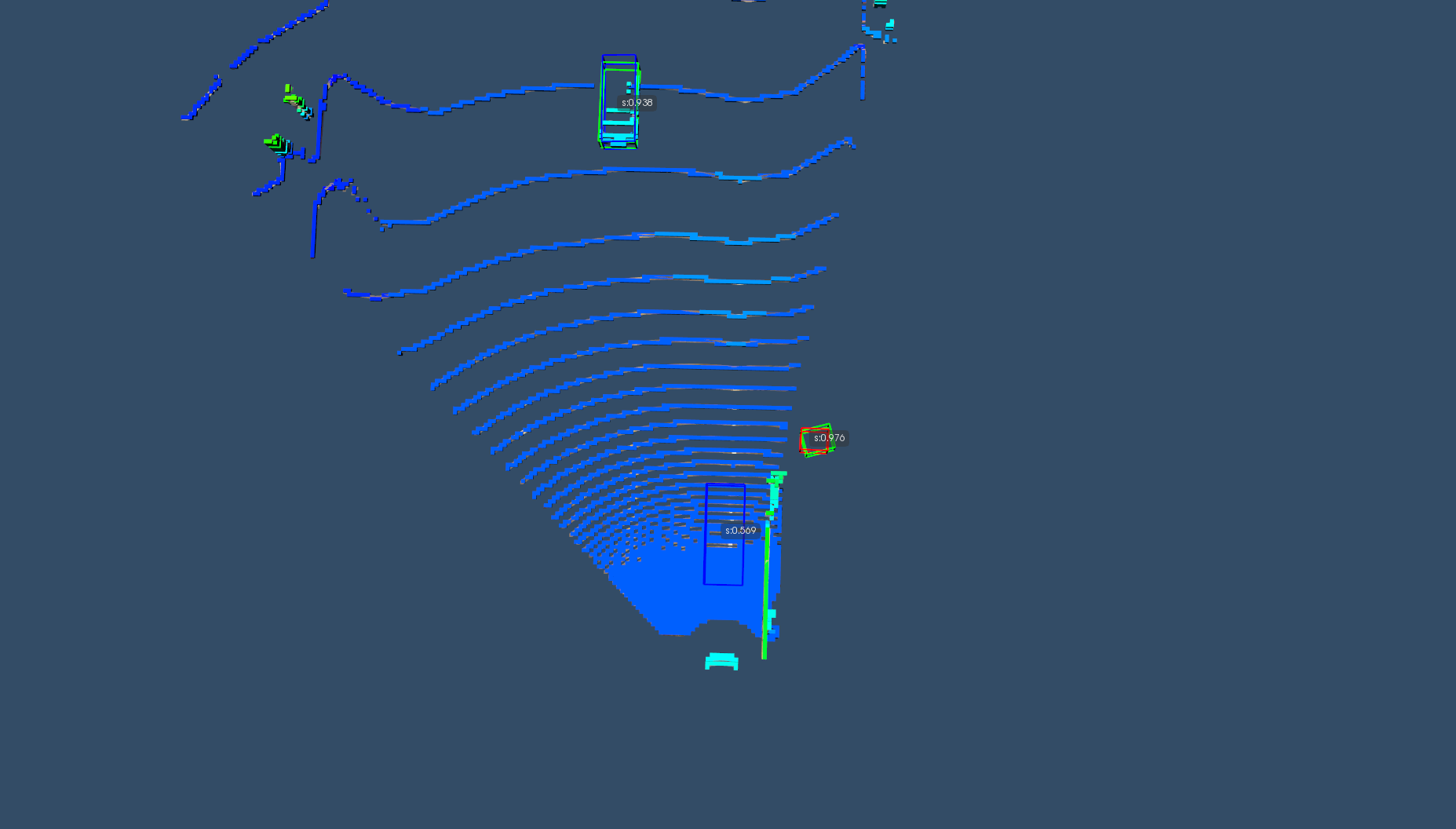} \includegraphics[width=0.23\textwidth]{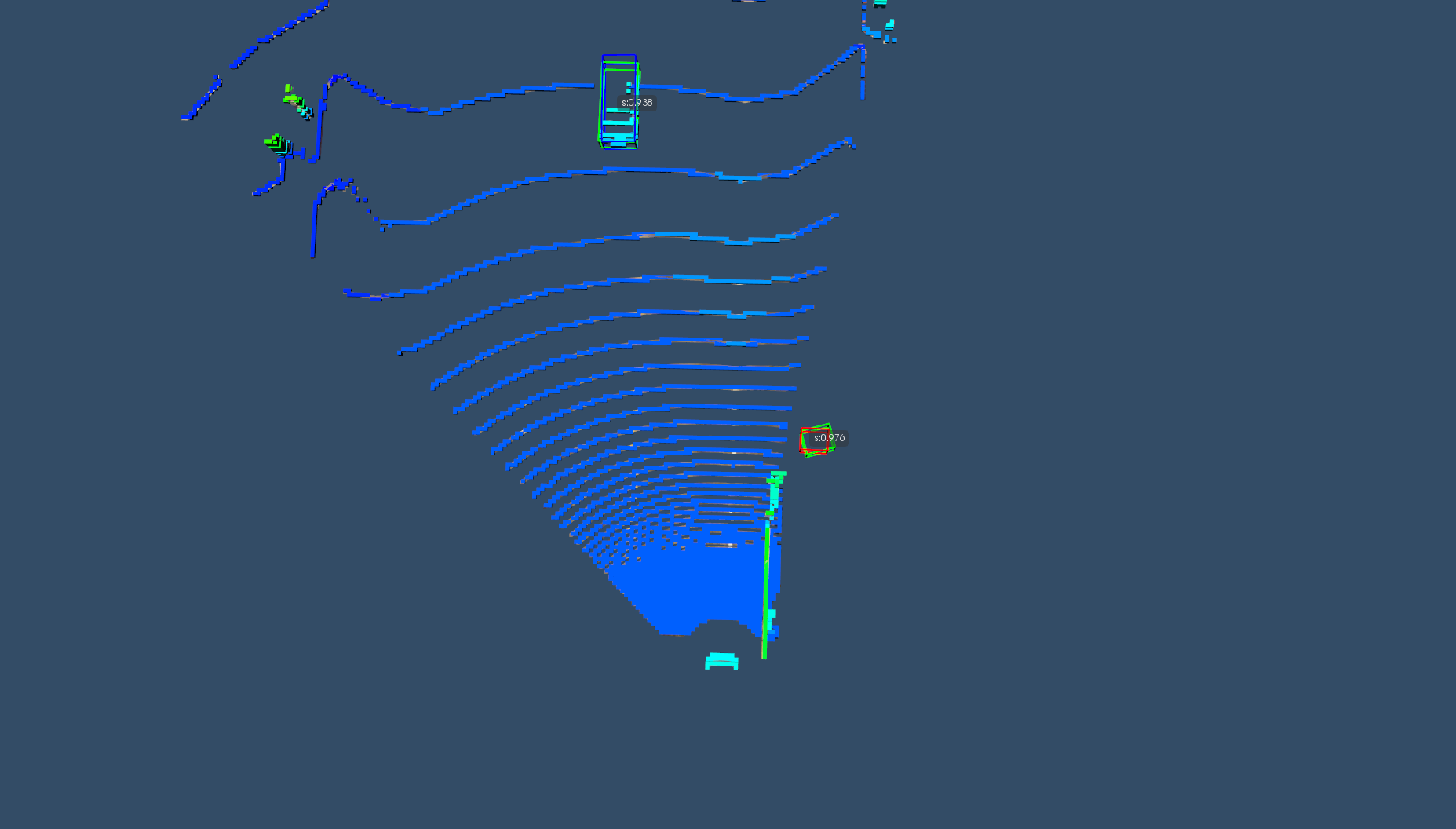}\\
  \caption{Top: Ego-vehicle image and point cloud of scenario 1. Middle: Ego-vehicle (left) and oncoming vehicle detections (right). Oncoming vehicle has inserted a false detection in front of the ego-vehicle. Bottom: Cooperative perception results without (left) and with (right) the plausibility checker. Ground truth is shown in green, detection score on boxes.}
  \label{trust_scenario_ego_persp}
  \vspace{-3mm}
\end{figure}

The TruPercept cooperative message evaluation mechanism is able to correctly detect the pedestrian which is occluded from the ego-vehicle perspective. Unfortunately the false detection, although the score is decreased, is still present. However, if the plausibility checker is also run, it can eliminate the possibility that the false detection exists while still maintaining the pedestrian which is occluded. The image and a top-down view of the LiDAR point cloud, obtained from the ego-vehicle perspective, is displayed in Figure \ref{trust_scenario_ego_persp} along with detections from each car and final outputs. The false detection is completely removed and the occluded pedestrian is detected.

\subsection{Scenario 2: Vehicle Trust}
\label{vehicleTrustScenario}
This experiment attempts to show that the TruPercept model can correctly identify a malicious vehicle. Trust is accrued over time by the evaluations of peers. To mimic this, the single frame from the previous scenario is replicated 100 times and the trust score is examined at various instances in time. The false detections are inserted after 50 frames so the vehicle has time to build up trust. A freshness timeout $f = 50$ frames is used so that any evaluation further than 50 frames in the past is discarded and not used towards the vehicle trust. A secondary situation is introduced where the oncoming vehicle places three false detections to compare when there is more false information.




The trust for the ego-vehicle remains at 0.82. For the malicious vehicle, the trust value begins at 0.81 and decreases to 0.15 by frame 100. When three false vehicles are added the trust value decreases faster and balances out at zero. Trustworthy vehicles remain unchanged, thus the TruPercept algorithm was able to correctly identify the malicious vehicle. Note that there is a third vehicle not depicted in the scene images which has a clear view of all objects in the scene but is further away. It contributes to the positive detection of the occluded pedestrian and has a trust score of 0.97.

\section{Quantitative Experiments}
\label{baselineExperiment}

This analysis shows quantifiable effects of using cooperative perception on the general perceptual pipeline. The TruPercept dataset is, to the best of our knowledge, the first of its kind, and enables cooperative perception model evaluation on regular driving data. The KITTI evaluation criteria of easy, moderate, and hard are defined by limiting values on occlusion, truncation, and minimum height (in pixels) of objects. This scope is too narrow, as cooperative perception allows detection of objects which are completely occluded, truncated, or are far in the distance. To provide a deeper inspection, the `All' criteria, which encompasses any object within the forward facing 90 degree FoV area up to 140 metres away, is also evaluated.

\subsection{Cooperative vs. Local Perception}
The first experiment is designed to evaluate if the TruPercept model improves upon single perspective methods. AVOD \cite{avod}, a state-of-the-art 3D object detector, is trained using the PreSIL dataset \cite{presil} and the mean AP score on the TruPercept data is used as a baseline. Next, AVOD is run for every nearby vehicle (within a 100m radius), and the TruPercept model is run to obtain updated scores. The scores of these methods are displayed in Table \ref{table_results_trupercept}. Several variations of TruPercept are
\begin{enumerate*}
    \item Weighted Average, $\eta = 0$
    \item Summed Positive and Negative, $\eta = -1$
    \item Local detections were used primarily. Any non-matching TruPercept 2 detection $\theta$ is used if, from the ego-vehicle perspective $v$, $\Phi^v(\theta) < 10$ and the plausibility check returns true.
\end{enumerate*}
For shown results $\tau = 0.1$, $\gamma_l = 0$, $\gamma_u = 100$ for cars, and $40$ for pedestrians.

\begin{table}[htbp]
\caption{Results: TruPercept (\% AP)}\label{table_results_trupercept}
\vspace{-3mm}
\begin{center}
\setlength\tabcolsep{2pt}
\begin{tabular}{|l|l|l|l|l|l|l|l|l|}
\hline
 & \multicolumn{4}{c|}{Car} & \multicolumn{4}{c|}{Pedestrian}\\
\hline
Method & Easy & Mod. & Hard & All & Easy & Mod. & Hard & All\\
\hline
AVOD & \cellcolor{blue!25}66.8 & \cellcolor{blue!25}62.8 & \cellcolor{blue!25}52.1 & 36.3 & \cellcolor{blue!25}84.2 & 78.7 & 75.1 & 57.3 \\
\hline
TruPercept 1 & 54.6 & 48.9 & 43.9 & 31.5 & 76.7 & 68.7 & 70.0 & 54.0 \\
\hline
TruPercept 2 & 57.2 & 51.4 & 46.0 & 32.8 & 78.6 & 70.4 & 71.4 & 55.0 \\
\hline
TruPercept 3 & 64.8 & 62.1 & 51.1 & \cellcolor{blue!25}36.7 & 83.4 & \cellcolor{blue!25}78.8 & \cellcolor{blue!25}75.8 & \cellcolor{blue!25}58.5 \\
\hline
\end{tabular}
\end{center}
\vspace{-3mm}
\end{table}

The most important insight is that none of the tested methods improve the perceptual accuracy by a meaningful margin over the local perception methods. This shows the importance of testing any modules which are added to the perceptual track with the entire system instead of only scenarios. Two plausible explanations for the poor TruPercept results are:
\begin{enumerate*}
    \item The alignment issues with the synthetic data could reduce performance for cooperative, but not local, perception.
    \item Each perspective can introduce false detections, when aggregating between multiple vehicles this could result in more false positives. By investigating the precision-recall curves for several of the models it is evident that for many of the poorly performing cooperative perception models the precision is much lower. This is especially true for lower recall values, likely caused by false detections from nearby vehicles being inserted with high scores.
\end{enumerate*}


One point to note is that pedestrians have higher detection scores than cars, a reversal from typical real-world data. This could be from pedestrian bounding boxes having identical sizes in the PreSIL data, making regression easier. The PreSIL data also has larger quantities of vehicles such as trucks, busses, and SUVs which are not part of the `Car' class. This could cause confusion for a detection network which is only attempting to label cars, resulting in more false positives. This could be rectified by training a network to detect all classes.

\subsection{Position and Orientation}
The hypothesis is that vehicles which are closer to an object will provide more accurate positioning and orientation information. This could lead to increased accuracy if objects which are detected are not meeting the KITTI standard of 0.5 IoU to be a true positive. For this experiment, the local AVOD detections have the position and orientation set to that of the closest detecting vehicle in each matching detections list. Results in Table \ref{table_position_orientation} show an increase in AP of up to 5\%. It is revealing that the largest improvement to local perception is gained simply by modifying the position and orientation of local detections. This enables the model to ignore false positives from nearby vehicles and yet still improve the accuracy of local detections.

\begin{table}[htbp]
\caption{Results: Updated position/orientation (\% AP)}\label{table_position_orientation}
\begin{center}
\vspace{-3mm}
\setlength\tabcolsep{2pt}
\begin{tabular}{|l|l|l|l|l|l|l|l|l|}
\hline
 & \multicolumn{4}{c|}{Car} & \multicolumn{4}{c|}{Pedestrian}\\
\hline
Method & Easy & Mod. & Hard & All & Easy & Mod. & Hard & All\\
\hline
AVOD & 66.8 & \cellcolor{blue!25}62.8 & 52.1 & 36.3 & 84.2 & 78.7 & 75.1 & 57.3 \\
\hline
Corrected & \cellcolor{blue!25}69.5 & 61.0 & \cellcolor{blue!25}54.1 & \cellcolor{blue!25}37.8 & \cellcolor{blue!25}89.0 & \cellcolor{blue!25}80.6 & \cellcolor{blue!25}80.1 & \cellcolor{blue!25}61.0 \\
\hline
\end{tabular}
\end{center}
\end{table}
\vspace{-5mm}

\subsection{Trust Levels}
The previous experiments evaluated detections from honest (although potentially not correct) vehicles. This experiment augments the dataset with malicious and unreliable vehicles to evaluate the performance of the TruPercept models, while being exposed to entities with non-optimal behaviours.
\begin{itemize}
    \item Trustworthy: detections are taken from the output of the local detection network. Results in Table \ref{table_results_trupercept}.
    \item Unreliable: for each local detection, 10\% chance to remove, 10\% add a randomly positioned detection.
    \item Malicious: a vehicle and pedestrian are inserted in front of the ego-vehicle for every frame from 10\% of vehicles.
\end{itemize}


Full results are available with the dataset. The car class shows a significant performance reduction (~10\%) when introducing the unreliable behaviours. Strangely, the pedestrian class actually improves in all categories for the TruPercept models. If the unreliable vehicles had bad detections to begin with, the final score of these detections could be reduced. Furthermore, false detections also have a chance to be removed, these could be harder to identify as false detections since they will be located where points are, instead of random locations which could contain no points. The malicious detections decrease the performance significantly, even for pedestrians. The malicious detections behaviour represents a coordinated attack between 10\% of the vehicles specifically targeted towards the ego-vehicle. This shows a weakness in the model towards coordinated attacks.


The mean trust values at the termination of the experiments were: 0.27 trustworthy, 0.25 unreliable, and 0.13 malicious. The trust value for unreliable vehicles is not significantly different than trustworthy vehicles. It is likely that behaviours such as the unreliable behaviour, which mostly present true detections, will be able to fool the current system. The trust model was able to detect blatantly malicious behaviour with a much higher success. Average trust values are fairly low, likely due to the amount of false positives which are being forwarded. If local detection accuracy increases, trust values would be expected to increase. These behaviours are simply a glimpse at strategies malicious entities could employ. In the future, great effort should be placed on creating and testing behaviours which could precipitate dangerous situations.

\section{Discussion and Conclusion}
In the future, we can imagine expanding our trust modeling, inspired by the work of VANETs researchers, such as supporting role-based trust modeling (e.g., police cars with greater initial trustworthiness) (as in \cite{relaycontrol}, \cite{MinhasIncentives2}) or adding incentives for honest reporting (\cite{MinhasIncentives1}, \cite{MinhasIncentives2}). Zhang et al. \cite{relaycontrol} also suggest that clustering methods may help to reduce false information spread. We note that since objects are naturally geographically clustered, relay control groups come to mind; by integrating this into the TruPercept solution, information could perhaps be drastically reduced (e.g., one score per matching object group). This is an avenue for future work.

Other researchers have new ideas for the use of a central server to deal with malicious entities (e.g., Li et al.'s \cite{rba} reputation-based announcement system which aggregates reports by reputation score). This may suggest an expanded role for our central system. More sophisticated trust modelling using Bayesian methods and Partially Observable Markov Decision Processes (POMDPs) are another area of study (e.g., Raya's system \cite{raya} which allows not knowing about events, Zhang et al. \cite{pomdp}'s use of partially observable Markov decision processes to enable dynamic updates). One way in which we could try for expanded trust modelling would be considering each new object in our framework with a POMDP. Allowing more uncertainty about vehicles is advocated by Balkrishan \cite{Balakrishnan} though their context is MANETs, while Souissi et al. \cite{vanetClasses} suggest more potential by examining similarity of messages received from different sources, which may help to determine the risk of believing a given report. These are other possible directions for richer trust modelling components.

In conclusion, multiple trust modelling methods for ad hoc networks were presented, with a majority being used for VANETs. The topic was evaluated from a new perspective: integration with 3DOD. A multi-agent solution attempting to increase 3DOD range and accuracy for AVs was presented. The TruPercept dataset, including all trust scenarios, as well as all tools, evaluation, and model code are publicly available at \url{https://tinyurl.com/y2nwy52o}.

Four broader-scoped takeaways are:
\begin{enumerate*}
    \item Cooperative perception can be used to more accurately localize objects.
    \item Cooperative perception creates a gargantuan security flaw for AVs since the correctness of incoming information cannot be guaranteed.
    \item Cooperative perception techniques need to incorporate trust modelling or incoming information evaluation. It is fairly easy to concoct situations which could cause an AV to act erratically and create dangerous situations.
    \item Cooperative perception methods should be  integrated into perceptual pipelines then evaluated on large datasets. Creating a module to prevent a contrived scenario from occurring is not necessarily hard. Integrating modules into complex perceptual systems to improve overall performance and prevent a plethora of prospectively hazardous scenarios is a more onerous task.
\end{enumerate*}




\bibliographystyle{IEEEtran}
\bibliography{IEEEabrv,mybib}

\begin{thebibliography}{10}
\providecommand{\url}[1]{#1}
\csname url@rmstyle\endcsname
\providecommand{\newblock}{\relax}
\providecommand{\bibinfo}[2]{#2}
\providecommand\BIBentrySTDinterwordspacing{\spaceskip=0pt\relax}
\providecommand\BIBentryALTinterwordstretchfactor{4}
\providecommand\BIBentryALTinterwordspacing{\spaceskip=\fontdimen2\font plus
\BIBentryALTinterwordstretchfactor\fontdimen3\font minus
  \fontdimen4\font\relax}
\providecommand\BIBforeignlanguage[2]{{%
\expandafter\ifx\csname l@#1\endcsname\relax
\typeout{** WARNING: IEEEtran.bst: No hyphenation pattern has been}%
\typeout{** loaded for the language `#1'. Using the pattern for}%
\typeout{** the default language instead.}%
\else
\language=\csname l@#1\endcsname
\fi
#2}}

\bibitem{collaborativePerception}
S.~Khan, F.~Andert, N.~Wojke, J.~Schindler, A.~Correa, and A.~Wijbenga,
  ``Towards collaborative perception for automated vehicles in heterogeneous
  traffic,'' in \emph{Advanced Microsystems for Automotive Applications 2018},
  J.~Dubbert, B.~M{\"u}ller, and G.~Meyer, Eds.\hskip 1em plus 0.5em minus
  0.4em\relax Cham: Springer International Publishing, 2019, pp. 31--42.

\bibitem{MinhasIncentives1}
U.~F. {Minhas}, J.~{Zhang}, T.~{Tran}, and R.~{Cohen}, ``Intelligent agents in
  mobile vehicular ad-hoc networks: Leveraging trust modeling based on direct
  experience with incentives for honesty,'' in \emph{2010 IEEE/WIC/ACM
  International Conference on Web Intelligence and Intelligent Agent
  Technology}, vol.~2, Aug 2010, pp. 243--247.

\bibitem{ensemble}
T.~G. Dietterich, ``Ensemble methods in machine learning,'' in \emph{Multiple
  Classifier Systems}.\hskip 1em plus 0.5em minus 0.4em\relax Berlin,
  Heidelberg: Springer Berlin Heidelberg, 2000, pp. 1--15.

\bibitem{multiSensorDataFusionPlausibilityChecking}
M.~{Obst}, L.~{Hobert}, and P.~{Reisdorf}, ``Multi-sensor data fusion for
  checking plausibility of v2v communications by vision-based multiple-object
  tracking,'' in \emph{2014 IEEE Vehicular Networking Conference (VNC)}, Dec
  2014, pp. 143--150.

\bibitem{networkCollisions}
\BIBentryALTinterwordspacing
Q.~Xu, T.~Mak, J.~Ko, and R.~Sengupta, ``Vehicle-to-vehicle safety messaging in
  dsrc,'' in \emph{Proceedings of the 1st ACM International Workshop on
  Vehicular Ad Hoc Networks}, ser. VANET '04.\hskip 1em plus 0.5em minus
  0.4em\relax New York, NY, USA: ACM, 2004, pp. 19--28. [Online]. Available:
  \url{http://doi.acm.org/10.1145/1023875.1023879}
\BIBentrySTDinterwordspacing

\bibitem{reladec}
H.~Volos, T.~Bando, and K.~Konishi, ``Reladec: Reliable latency decision
  algorithm for connected vehicle applications,'' in \emph{IEEE Intelligent
  Vehicles Symposium}, 2019.

\bibitem{riedl}
K.~Riedl, S.~Kurscheid, A.~Noll, J.~Betz, and M.~Lienkamp, ``Road network
  coverage models for cloud-based automotive applications: A case study in the
  city of munich,'' in \emph{IEEE Intelligent Vehicles Symposium}, 2019.

\bibitem{correa}
A.~Correa, R.~Alms, J.~Gozalvez, M.~Sepulcre, R.~B. Michele~Rondinone,
  L.~Lücken, and G.~Thandavarayan, ``Infrastructure support for cooperative
  maneuvers in connected and automated driving,'' in \emph{IEEE Intelligent
  Vehicles Symposium}, 2019.

\bibitem{KimMultiVehicleSLAM}
S.~{Kim}, Z.~J. {Chong}, B.~{Qin}, X.~{Shen}, Z.~{Cheng}, W.~{Liu}, and M.~H.
  {Ang}, ``Cooperative perception for autonomous vehicle control on the road:
  Motivation and experimental results,'' in \emph{2013 IEEE/RSJ International
  Conference on Intelligent Robots and Systems}, Nov 2013, pp. 5059--5066.

\bibitem{JimenezMultiVehicleSLAM}
\BIBentryALTinterwordspacing
A.~Jiménez-González, J.~R. Martínez-de Dios, and A.~Ollero, ``An integrated
  testbed for cooperative perception with heterogeneous mobile and static
  sensors,'' \emph{Sensors}, vol.~11, no.~12, p. 11516–11543, Dec 2011.
  [Online]. Available: \url{http://dx.doi.org/10.3390/s111211516}
\BIBentrySTDinterwordspacing

\bibitem{LiCooperativePerception}
H.~Li, ``Cooperative perception : Application in the context of outdoor
  intelligent vehicle systems,'' Ph.D. dissertation, l’École Nationale
  Supérieure des Mines de Paris, Sept 2012.

\bibitem{allig}
C.~Allig and G.~Wanielik, ``Alignment of perception information for cooperative
  perception,'' in \emph{IEEE Intelligent Vehicles Symposium}, 2019.

\bibitem{datasetSummary}
\BIBentryALTinterwordspacing
D.~Feng, C.~Haase{-}Schuetz, L.~Rosenbaum, H.~Hertlein, F.~Duffhauss,
  C.~Glaser, W.~Wiesbeck, and K.~Dietmayer, ``Deep multi-modal object detection
  and semantic segmentation for autonomous driving: Datasets, methods, and
  challenges,'' \emph{CoRR}, vol. abs/1902.07830, 2019. [Online]. Available:
  \url{http://arxiv.org/abs/1902.07830}
\BIBentrySTDinterwordspacing

\bibitem{arnold}
E.~Arnold, O.~Y. Al-Jarrah, M.~Dianati, S.~Fallah, D.~Oxtoby, and
  A.~Mouzakitis, ``Cooperative object classification for driving
  applications,'' in \emph{IEEE Intelligent Vehicles Symposium}, 2019.

\bibitem{cooper}
\BIBentryALTinterwordspacing
Q.~Chen, S.~Tang, Q.~Yang, and S.~Fu, ``Cooper: Cooperative perception for
  connected autonomous vehicles based on 3d point clouds,'' \emph{CoRR}, vol.
  abs/1905.05265, 2019. [Online]. Available:
  \url{http://arxiv.org/abs/1905.05265}
\BIBentrySTDinterwordspacing

\bibitem{cam}
``{Intelligent Transport Systems (ITS); Vehicular Communications; Basic Set of
  Applications; Part 2: Specification of Cooperative Awareness Basic
  Service},'' ETSI, Tech. Rep. EN 302 637-2 V1.3.2, Nov 2014.

\bibitem{cpm}
``{Intelligent Transport Systems (ITS); Vehicular Communications; Basic Set of
  Applications; Analysis of the Collective Perception Service (CPS) },'' ETSI,
  Tech. Rep. TR 103 562 0.0.15 Draft, Jan 2019.

\bibitem{v2xDiscussion}
L.~Hobert, A.~Festag, I.~Llatser, L.~Altomare, F.~Visintainer, and A.~Kovacs,
  ``Enhancements of v2x communication in support of cooperative autonomous
  driving,'' \emph{IEEE Communications Magazine}, vol.~53, pp. 64--70, 12 2015.

\bibitem{pomdp}
S.~Chen, A.~A. Irissappane, and J.~Zhang, ``Pomdp-based decision making for
  fast event handling in vanets,'' in \emph{AAAI}, 2018.

\bibitem{rba}
Q.~Li, A.~Malip, K.~M. Martin, S.~Ng, and J.~Zhang, ``A reputation-based
  announcement scheme for vanets,'' \emph{IEEE Transactions on Vehicular
  Technology}, vol.~61, no.~9, pp. 4095--4108, Nov 2012.

\bibitem{MessagePropagationAndEvaluation}
C.~{Chen}, J.~{Zhang}, R.~{Cohen}, and P.~{Ho}, ``A trust modeling framework
  for message propagation and evaluation in vanets,'' in \emph{2010 2nd
  International Conference on Information Technology Convergence and Services},
  Aug 2010, pp. 1--8.

\bibitem{distributedTrustInOpenMultiAgentSystems}
\BIBentryALTinterwordspacing
Y.~Mass and O.~Shehory, ``Distributed trust in open multi-agent systems,'' in
  \emph{Proceedings of the Workshop on Deception, Fraud, and Trust in Agent
  Societies Held During the Autonomous Agents Conference: Trust in
  Cyber-societies, Integrating the Human and Artificial Perspectives}.\hskip
  1em plus 0.5em minus 0.4em\relax London, UK, UK: Springer-Verlag, 2001, pp.
  159--174. [Online]. Available:
  \url{http://dl.acm.org/citation.cfm?id=646674.701823}
\BIBentrySTDinterwordspacing

\bibitem{avod}
J.~Ku, M.~Mozifian, J.~Lee, A.~Harakeh, and S.~Waslander, ``Joint 3d proposal
  generation and object detection from view aggregation,'' \emph{IROS}, 2018.

\bibitem{fpointnet}
\BIBentryALTinterwordspacing
C.~R. Qi, W.~Liu, C.~Wu, H.~Su, and L.~J. Guibas, ``Frustum pointnets for 3d
  object detection from {RGB-D} data,'' \emph{CoRR}, vol. abs/1711.08488, 2017.
  [Online]. Available: \url{http://arxiv.org/abs/1711.08488}
\BIBentrySTDinterwordspacing

\bibitem{bayesOD}
\BIBentryALTinterwordspacing
A.~Harakeh, M.~Smart, and S.~L. Waslander, ``Bayesod: {A} bayesian approach for
  uncertainty estimation in deep object detectors,'' \emph{CoRR}, vol.
  abs/1903.03838, 2019. [Online]. Available:
  \url{http://arxiv.org/abs/1903.03838}
\BIBentrySTDinterwordspacing

\bibitem{omnet}
A.~Varga, ``The omnet++ discrete event simulation system,'' \emph{Proc.
  ESM'2001}, vol.~9, Jan 2001.

\bibitem{sumo}
M.~Behrisch, L.~Bieker-Walz, J.~Erdmann, and D.~Krajzewicz, ``Sumo –
  simulation of urban mobility: An overview,'' \emph{Proceedings of SIMUL},
  vol. 2011, 10 2011.

\bibitem{carla}
\BIBentryALTinterwordspacing
A.~Dosovitskiy, G.~Ros, F.~Codevilla, A.~L{\'{o}}pez, and V.~Koltun, ``{CARLA:}
  an open urban driving simulator,'' \emph{CoRR}, vol. abs/1711.03938, 2017.
  [Online]. Available: \url{http://arxiv.org/abs/1711.03938}
\BIBentrySTDinterwordspacing

\bibitem{driveInMatrix}
\BIBentryALTinterwordspacing
M.~Johnson{-}Roberson, C.~Barto, R.~Mehta, S.~N. Sridhar, and R.~Vasudevan,
  ``Driving in the matrix: Can virtual worlds replace human-generated
  annotations for real world tasks?'' \emph{CoRR}, vol. abs/1610.01983, 2016.
  [Online]. Available: \url{http://arxiv.org/abs/1610.01983}
\BIBentrySTDinterwordspacing

\bibitem{gtavlidar}
\BIBentryALTinterwordspacing
X.~Yue, B.~Wu, S.~A. Seshia, K.~Keutzer, and A.~L. Sangiovanni{-}Vincentelli,
  ``A lidar point cloud generator: from a virtual world to autonomous
  driving,'' \emph{CoRR}, vol. abs/1804.00103, 2018. [Online]. Available:
  \url{http://arxiv.org/abs/1804.00103}
\BIBentrySTDinterwordspacing

\bibitem{presil}
B.~Hurl, K.~Czarnecki, and S.~Waslander, ``Precise synthetic image and lidar
  (presil) dataset for autonomous vehicle perception,'' in \emph{IEEE
  Intelligent Vehicles Symposium}, 2019.

\bibitem{relaycontrol}
\BIBentryALTinterwordspacing
J.~Zhang, C.~Chen, and R.~Cohen, ``Trust modeling for message relay control and
  local action decision making in vanets,'' \emph{Security and Communication
  Networks}, vol.~6, no.~1, pp. 1--14, 2012. [Online]. Available:
  \url{https://onlinelibrary.wiley.com/doi/abs/10.1002/sec.519}
\BIBentrySTDinterwordspacing

\bibitem{MinhasIncentives2}
U.~F. {Minhas}, J.~{Zhang}, T.~{Tran}, and R.~{Cohen}, ``A multifaceted
  approach to modeling agent trust for effective communication in the
  application of mobile ad hoc vehicular networks,'' \emph{IEEE Transactions on
  Systems, Man, and Cybernetics, Part C (Applications and Reviews)}, vol.~41,
  no.~3, pp. 407--420, May 2011.

\bibitem{raya}
M.~Raya, P.~Papadimitratos, V.~D. Gligor, and J.~. Hubaux, ``On data-centric
  trust establishment in ephemeral ad hoc networks,'' in \emph{IEEE INFOCOM
  2008 - The 27th Conference on Computer Communications}, April 2008, pp.
  1238--1246.

\bibitem{Balakrishnan}
\BIBentryALTinterwordspacing
V.~Balakrishnan, V.~Varadharajan, and U.~Tupakula, ``Subjective logic based
  trust model for mobile ad hoc networks,'' in \emph{Proceedings of the 4th
  International Conference on Security and Privacy in Communication Netowrks},
  ser. SecureComm '08.\hskip 1em plus 0.5em minus 0.4em\relax New York, NY,
  USA: ACM, 2008, pp. 30:1--30:11. [Online]. Available:
  \url{http://doi.acm.org/10.1145/1460877.1460916}
\BIBentrySTDinterwordspacing

\bibitem{vanetClasses}
I.~Souissi, N.~B. Azzouna, and T.~Berradia, ``Towards a self-adaptive trust
  management model for vanets,'' in \emph{Proceedings of the 14th International
  Joint Conference on e-Business and Telecommunications - Volume 6: SECRYPT,
  (ICETE 2017)}, INSTICC.\hskip 1em plus 0.5em minus 0.4em\relax SciTePress,
  2017, pp. 513--518.

\end{thebibliography}

\end{document}